\documentclass[twocolumn,floatfix,prl,aps,showpacs]{revtex4-2}
\usepackage{graphicx,amsmath,amssymb,color}
\usepackage{nicefrac}
\usepackage[titletoc,title]{appendix}
\usepackage[colorlinks,bookmarks=true,citecolor=blue,linkcolor=red,urlcolor=blue]{hyperref}

\newcommand{\be}{\begin{equation}}
\newcommand{\ee}{\end{equation}}

\newcommand{\ba}{\begin{eqnarray}}
\newcommand{\ea}{\end{eqnarray}}

\begin{document}

\title{Density dependent gauge field inducing emergent SSH physics, solitons and condensates in a discrete nonlinear Schr\"odinger equation}
\author{W. N. Faugno$^1$, Mario Salerno$^2$, and Tomoki Ozawa$^1$}
\affiliation{$^1$Advanced Institute for Materials Research (WPI-AIMR), Tohoku University, Sendai 980-8577, Japan}
\affiliation{$^2$Dipartimento di Fisica “E.R. Caianiello,” CNISM and Istituto Nazionale di Fisica Nucleare—Gruppo Collegato di Salerno, Universita d\'i
Salerno, Via Giovanni Paolo II, 84084 Fisciano (SA), Italy}
\date{\today}

\begin{abstract}
We investigate a discrete non-linear Schr\"odinger equation with dynamical, density-difference-dependent, gauge fields. We find a ground-state transition from a plane wave condensate to a localized soliton state as the gauge coupling is varied. Interestingly we find a regime in which the condensate and soliton are both stable. We identify an emergent chiral symmetry, which leads to the existence of a symmetry protected zero energy edge mode. The emergent chiral symmetry relates low and high energy solitons. These states indicate that the interaction acts both repulsively and attractively.
\end{abstract}

\maketitle

\textit{Introduction -}
The non-linear Schr\"odinger equation (NLSE) is an essential model for a variety of physical systems~\cite{Scott03,Malomed05}. In ultracold atomic physics, the NLSE describes Bose-Einstein condensates (BECs) in the dilute limit where only the leading order interaction is relevant~\cite{Pitaevskii03,Ablowitz04,Galati13,Liu21}. In optical systems, the NLSE describes light propagating in nonlinear media, originally employed to describe self-focusing beams of light~\cite{Chiao64,Ruan03,Yang10}. Since these initial investigations a wide range of studies into the NLSE and its extensions have been developed, identifying a variety of non-linear phenomena~\cite{Kaup78,Kenkre86,Vengalattore08,Ozawa13,Salerno15,Martone16,Deffner22}. Additionally, extensions have been made to simulate gauge theories \cite{Mishra16,ValentiRojas20,Bonkhoff21,Frolian22,Roell23}. Much work has gone into understanding the stability of NLSE solutions. This is in part because the onset of instability can often be associated with a phase transition to a stable non-dispersive bound state, termed a soliton~\cite{Baizakov02,Kevrekidis04,Carr04,Cornish06,Nath09,Malomed22}.

Solitons and other localized states have been predicted and verified in the NLSE and its extensions~\cite{Carr00,Carr00b,Salerno05,Abdullaev10}. These bound states result from the competition between the non-linearity and the kinetic dispersion. In the case of the discrete NLSEs \cite{Kevrekidis09,Kevrekidis01}, the existence of bound states termed discrete breathers has been shown to be generic~\cite{Flach08}. Such self-stabilizing solutions are widely studied for both their mathematical properties as well as their potential applications in developing communications technology. Recently, matter waves breathers have been  experimentally observed  by quenching the strength of the attractive interactions in a quasi one-dimensional Bose-Einstein condensate \cite{Luo20}. Engineering and manipulation of soliton modes has been achieved through various methods, including time-dependent modulation and spatial modulation \cite{Gouveia89,Kevrekidis05,Salerno08,Cai14,Rong21}.   

An interesting avenue explored in recent years has been to consider NLSE which contains a \textit{dynamical, density-dependent, gauge field}. Here, in the context of discrete (lattice) models, the gauge field is a nontrivial phase factor one has upon hopping between adjacent sites, and density-dependent gauge field means that the phase factor depends on the density of sites involved in the hopping \cite{Keilmann11,Rapp12,Greschner14,Greschner15,Meinert16,Wang20}. It has been shown that density-dependent hopping induced by sinusoidal temporal modulations of the nonlinearity in the discrete NLSE leads to the formation of compactons, i.e., nonlinear bound states that, unlike breathers, do not have exponential decay tails \cite{Abdullaev10}.
Recently, two of us proposed a novel type of density-dependent gauge field, in the context of interacting non-Hermitian quantum physics, where the hopping depends on the \textit{difference} of the density of sites involved in hopping \cite{Faugno22}; such a gauge field has a unique feature that the sign of the interaction is not determined \textit{a priori} but dynamically depends on the density gradient of the wavefunction profile. In this paper, we find that the NLSE derived from such a density-difference-dependent gauge field results in a variety of unique phenomena, including the coexistence of plane-wave condensate state and soliton state, the coexistence of attractive and repulsive solitons, and the emergent SSH-like physics with a symmetry-protected zero-energy mode. Our paper opens an avenue to explore nonlinear physics of this exotic form of gauge field.

\textit{The Discrete Density Difference Dependent Non-Linear Schrodinger Equation (D$^4$NLSE) -}
We consider a one dimensional discrete NLSE derived from the following classical Hamiltonian:
\begin{equation}
    H = \sum_j \Psi^*_{j+1}\bigg[-J+\gamma(n_{j+1}-n_j)\bigg]\Psi_j + c.c.,
    \label{eq:H}
\end{equation}
where $\Psi_j$ is the complex-valued amplitude of the wavefunction at site $j$, $n_j = \Psi_j^* \Psi_j$ is the corresponding density, $J$ is the density-independent part of hopping between adjacent sites, $\gamma$ is the coupling constant of the density-difference-dependent hopping which can generally be a complex number. We assume that the wavefunction is normalized as $\sum_j |\Psi_j|^2 = 1$. A crucial aspect of this Hamiltonian is that the sign of the interaction, either repulsive or attractive, is not determined \textit{a priori}, but rather determined dynamically through the density profile of the wavefuntion.

The corresponding NLSE can be obtained by computing the Poisson bracket $i\frac{d\Psi_j}{dt} = \left\{ \Psi_j, H\right\} = \frac{\delta H}{\delta \Psi_j^*}$:
\begin{align}
    i\frac{d\Psi_j}{dt} &= \left\{ -J + \gamma (2n_j - n_{j-1}) \right)\} \Psi_{j-1}
    -\gamma \Psi_j^2 \Psi_{j+1}^* \notag \\
    &+\left\{ -J - \gamma^* (2n_j - n_{j+1}) \right)\} \Psi_{j+1} + \gamma^* \Psi_j^2 \Psi_{j-1}^*,
\label{D4NLSE}
\end{align}
which is the one-dimensional Discrete Density Difference Dependent (D$^4$) Non-linear Schr\"odinger equation we are going to explore in this paper.

We note that the Hamiltonian in Eq.~\ref{eq:H} can be also thought of as quantum Hamiltonian by identifying $\Psi^*_j$ and $\Psi_j$ as creation and annihilation operators of a particle at site $j$. Then the Poisson bracket mentioned above should be replaced by a commutator, and D$^4$ NLSE describes a situation where the mean-field theory is applicable and the Gross-Pitaevskii treatment of the condensate wavefunction is justified.

\textit{Ground state phase diagram:---}
We explore the stationary state of the D$^4$ NLSE evolving in time as $\Psi_j (t) = e^{-i\mu t}\Psi_j(0)$, where $\mu$ is the chemical potential. Of particular interest is the ground state, which is the state with the lowest energy. We can numerically explore the ground state by the method of imaginary-time propagation, which is to start from a random initial state and simulate the time evolution of the D$^4$ NLSE with an imaginary time, that is, to consider the equation obtained after setting $t = -i\tau$ and simulate the evolution with respect to $\tau$. After evolving long enough time in $\tau$, one converges to the ground state (if the ground state is unique) \cite{Antoine15,Antoine17}. Assuming periodic boundary condition and applying imaginary time evolution method to various values of $\gamma$, we see that the ground state is either a plane-wave condensate state or a sharply localized soliton-like state, which we explain now.

We first note that, for states with a uniform density, $n_j = n_{j+1}$, $\gamma$-dependence in the Hamiltonian vanishes. The plane-wave state $e^{ikx}$ with $k = 2\pi (\mathrm{integer})/L$, where $L$ is the number of lattice sites, is thus a valid stationary state with chemical potential $\mu = -2J \cos (k)$. Among the plane wave solutions, the ground state is the $k = 0$ state, and the corresponding chemical potential as well as the energy is $\mu = E = -2J$.

Now the question to ask is if one can obtain a state with lower energy by allowing the density to vary. 
We indeed find that the system can host localized stationary states which we call solitons. Essential features of the soliton can be captured through the following ansatz where the soliton is spread over only three sites around a site $n$:
\begin{align}
        \Psi_{n\pm 1} &= e^{i\phi_\pm}\sqrt{(1-\alpha)/2}, & 
        \Psi_n &= \sqrt{\alpha},
\end{align}
where $\phi_-, \phi_+, \alpha$ are variational parameters which we take as real numbers.
We expect that the interaction energy dominates in such soliton states.
The transition between the $k=0$ plane wave and the localized ansatz can be solved analytically for purely real $\gamma$ and is predicted to occur at $|\gamma| = 2.53J$. For complex $\gamma$, the critical $|\gamma|$ weakly depends on its phase, decreasing slightly as the phase goes towards purely imaginary $\gamma$ where the transition occurs around $|\gamma| \approx 2.5J$. We find from imaginary time evolution that the transition occurs at $|\gamma| \approx 2.04J$ for fully real $\gamma$ and $|\gamma| \approx 1.9J$ for fully imaginary $\gamma$; these values are smaller than the prediction from the three-site ansatz we obtained above because the soliton state around the transition point can be more spread and/or asymmetric. We numerically confirm that, typically, more than 95\% of the weight of the soliton lies on three sites, and the three-site ansatz becomes more and more accurate as $|\gamma|$ becomes larger.

\begin{figure}
    \centering
    \includegraphics[width = 0.5\textwidth]{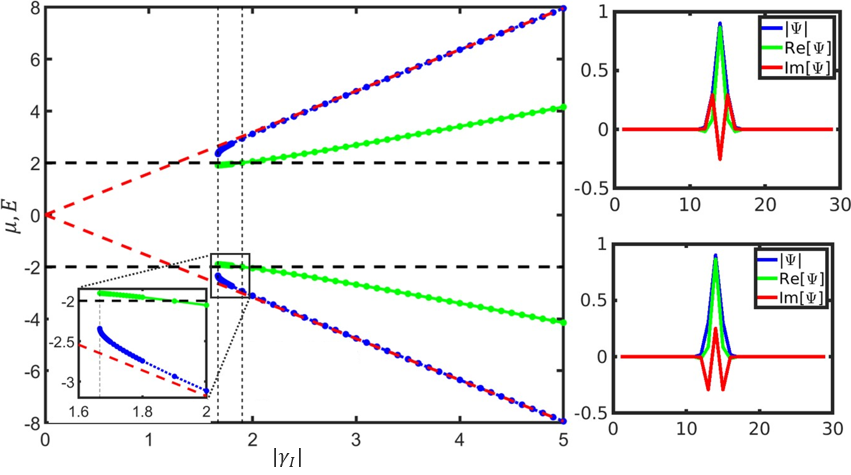}
    \caption{\textbf{Left Panel.} Energy (green solid curve) and the chemical potential (blue dotted curve) of the localized soliton state, in units of $J$, obtained from the imaginary time propagation method, as a function of the gauge coupling which we assume to be purely imaginary, $\gamma = i\gamma_I$.
    The red dashed lines correspond to the chemical potential of the localized states obtained when setting $J = 0$. The dashed horizontal black lines correspond to the lowest and highest plane wave condensate states at $\mu = E = \pm 2J$. The inset highlights the transition point and soliton disappearance. 
    \textbf{Right Panels.} Ground state and highest energy state wavefunction for $J=1$, $\gamma = -4.0i$ and length $L=29$. The blue, green and red curves are the wave function amplitude, real part and imaginary part, respectively.}
    \label{fig:GSEn_Sol}
\end{figure}

We provide a more detailed analysis for the case of purely imaginary $\gamma = i\gamma_I$. The results are qualitatively the same for any fixed phase of $\gamma$ with slight quantitative differences as mentioned above. In Fig.~\ref{fig:GSEn_Sol}, we plot the energy and the chemical potentials for the lowest and highest energy localized soliton states as a function of $|\gamma_I|$ obtained  from imaginary time evolution. These results were found in excellent agreement also with the ones obtained  by means of self-consistent numerical diagonalizations  of the nonlinear eigenvalue problem  associated with Eq.~\ref{D4NLSE} \cite{Salerno05}. The symmetry around $E = 0$ is a consequence of an emergent chiral symmetry which we discuss later. At $|\gamma_I|>1.9J$, the solitons are the lowest energy states. In the region $1.67J < |\gamma_I| < 1.9J$, the soliton solution still exists as a stationary state but its energy is higher than the energy of the plane wave condensate state. As $|\gamma_I|$ approaches $\approx 1.67J$ from above, the chemical potential of the soliton changes rapidly, and at $|\gamma_I| < 1.67J$, the soliton state no longer exists as a stationary state but rather merges with the extended modes. The energy and the chemical potential of the soliton state just above $|\gamma_I| \approx 1.67J$ are $E \approx -1.9J$ and $\mu \approx -2.3J$. Note that we have studied this model in the continuum and found that in solving for a localized state, the energy is unbounded from below, suggesting that soliton formation is a genuine lattice effect (see supplemental material).

\textit{Stability of the condensate:---}
We have just seen that the ground state changes from plane-wave condensate state to localized soliton state as one increases the coupling $|\gamma|$. The mechanism behind the formation of soliton ground state is qualitatively different from more conventional soliton formation in, for example, NLSE with a cubic nonlinearity, where the development of modulational instability of plane-wave condensate state leads to the formation of solitons \cite{Konotop02}. In contrast, in the D$^4$ NLSE, there can be regions in the parameter space where the zero momentum plane wave condensate and the soliton both exist as stable stationary states. Depending on the value of $\gamma$, the condensate can become unstable, as we show soon, but the instability of condensate and the formation of soliton appear to be not directly related. We now explore this linear stability of the condensate phase and present a complete phase diagram, Fig~\ref{fig:PhaseDiagram}, including the stability analysis. (In Appendix~\ref{app:ctnlim}, we also present the continuum model and demonstrate that its plane wave solutions have the same stability condition.)

\begin{figure}
    \centering
    \includegraphics[width=0.4\textwidth]{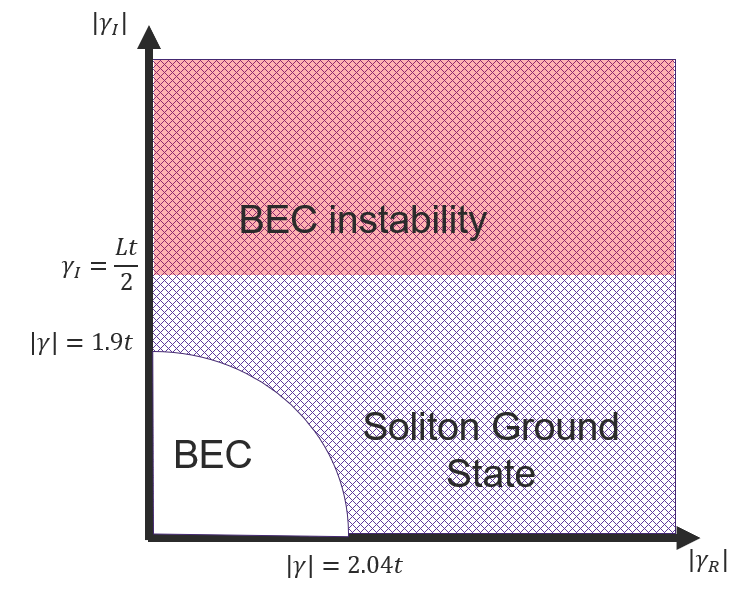}
    \caption{Phase Diagram of the D$^3$NLSE as a function of the complex gauge field coupling $\gamma$. The complex gauge coupling is defined as $\gamma_R + i\gamma_I$, $J$ is the single particle hopping parameter, and $L$ is the length of the chain. Note this phase diagram only considers the linear stability. Numerically we have observed that the instability can occur for completely real $\gamma$, likely due to higher order instabilities.}
    \label{fig:PhaseDiagram}
\end{figure}

To investigate the linear stability of a condensate upon small perturbation, we introduce the Fourier transformed Hamiltonian. Defining the wavefunction in momentum space with momentum $k$ by $\tilde{\Psi}_k$ as
\begin{equation}
    \Psi_j = \frac{1}{\sqrt{L}}\sum_k \tilde{\Psi}_ke^{-ijk}.
\end{equation}
the Hamiltonian is
\begin{align}
        &H = \sum_k -2J\cos{k}\tilde{\Psi}^*_k\tilde{\Psi}_k + \frac{1}{L} \sum_{k,k',q} V(q,k) \tilde{\Psi}^*_{k} \tilde{\Psi}^*_{k'} \tilde{\Psi}_{k+k'-q} \tilde{\Psi}_{q}\notag \\
        &V(q,k) = 2i\gamma_R(\sin{k} - \sin{q}) + 2i\gamma_I(\cos{q} - \cos{k}).
    \label{eq:Hk}
\end{align}
where we have introduced the notation $\gamma=\gamma_R + i\gamma_I$.
The time evolution equation in momentum space is then
\begin{align}
        &i\dot{\tilde{\Psi}}_k = -2J(\cos{k}-1)\tilde{\Psi}_k  \label{eq:TimeEvo} \\
        &+ \sum_{q,k'}V(q,k)\tilde{\Psi}^*_{k'}\tilde{\Psi}_{k+k'-q}\tilde{\Psi}_q + \sum_{q,k'}V(q,k')\tilde{\Psi}^*_{k'}\tilde{\Psi}_{k+k'-q}\tilde{\Psi}_q \notag
\end{align}
We assume that the condensate takes place at zero momentum, and add small perturbation at momentum $p$ to investigate its stability; we set $\tilde{\Psi}_0 \approx 1$, $\tilde{\Psi}_{\pm p} = \delta \tilde{\Psi}_{\pm p}$ which should be kept up to linear order, and $\tilde{\Psi}_{k} = 0$ otherwise.
Inserting them into Eq~\ref{eq:TimeEvo} and ignoring terms of higher order in $\delta \tilde{\Psi}_{\pm p}$, we obtain the equation for the time derivative of $\delta \tilde{\Psi}_p$ as
\begin{equation}
    i\frac{d \delta \tilde{\Psi}_p}{dt} = -2J(\cos{p}-1)\delta \tilde{\Psi}_p + \frac{4i\gamma_I}{L}(\cos{p}-1)\delta \tilde{\Psi}^*_{-p},
\end{equation}
and a similar equation for the time evolution of $\delta \tilde{\Psi}_{-p}$.
This set of equations can be diagonalized, in a manner analogous to the Bogoliubov transformation, to obtain the dispersion relation, which is
\begin{equation}
    \epsilon (p) = 2J\sqrt{\left(1 - \frac{4\gamma_I^2 }{J^2L^2}\right)}(1-\cos{p}).
\end{equation}
This dispersion relation contains unusual features. First, at small momentum $p$, the dispersion is \textit{quadratic} rather than more standard linear excitation spectrum in NLSE. The effect of the interaction $\gamma$ is to modify the overall multiplicative factor, which is analogous to modifying the \textit{effective mass} of a particle. The second point to notice is that only the imaginary part of $\gamma$ enters the dispersion relation. The real part does not affect the excitation properties of the zero momentum condensate phase. The third point is that the dispersion becomes imaginary, implying the appearance of modulational instability, above a threshold value of $\gamma_I > J L / 2$; notably the transition depends on the \text{length} of the system $L$. The threshold to the instability of this condensate is drawn in the phase diagram Fig~\ref{fig:PhaseDiagram}. We point out that the stability condition is solely determined by $\gamma_I$ whereas the transition from condensate ground state to soliton ground state depends on $|\gamma|$, implying that these two transitions occur from different mechanisms. (The stability condition for a condensate at arbitrary momentum $k$ is presented in Appendix \ref{app:arbk} derived from Bogoliubov analysis of a bosonic system.)

\textit{Chiral Symmetry and Zero Mode:---}
In Fig.~\ref{fig:GSEn_Sol}, we see that the figure is symmetric around $E = 0$. This symmetry is a consequence of the chiral symmetry present in the system. By examining the equation of motion, we see that for a stationary state $\{ \Psi_j \}$, we obtain a state with the opposite chemical potential by multiplying the state by $e^{i\pi j}$, i.e. a factor of $-1$ to every other site. In Fig.~\ref{fig:GSEn_Sol}, the wavefunctions of the ground state and the highest energy state are plotted for $\gamma = -4i$ to demonstrate that they are precisely related by the transformation $e^{i\pi j}$. Note that the existence of these two bound states is a signature of the attractive and repulsive nature of our interaction. Typically the appearance of low energy bound states is a signature of an attractive interaction \cite{Khaykovich02,Cornish06} whereas the existence of high energy bound states is a signature of a repulsive interaction \cite{Ishikawa80}.

\begin{figure}
    \centering
    \includegraphics[width = 0.4\textwidth]{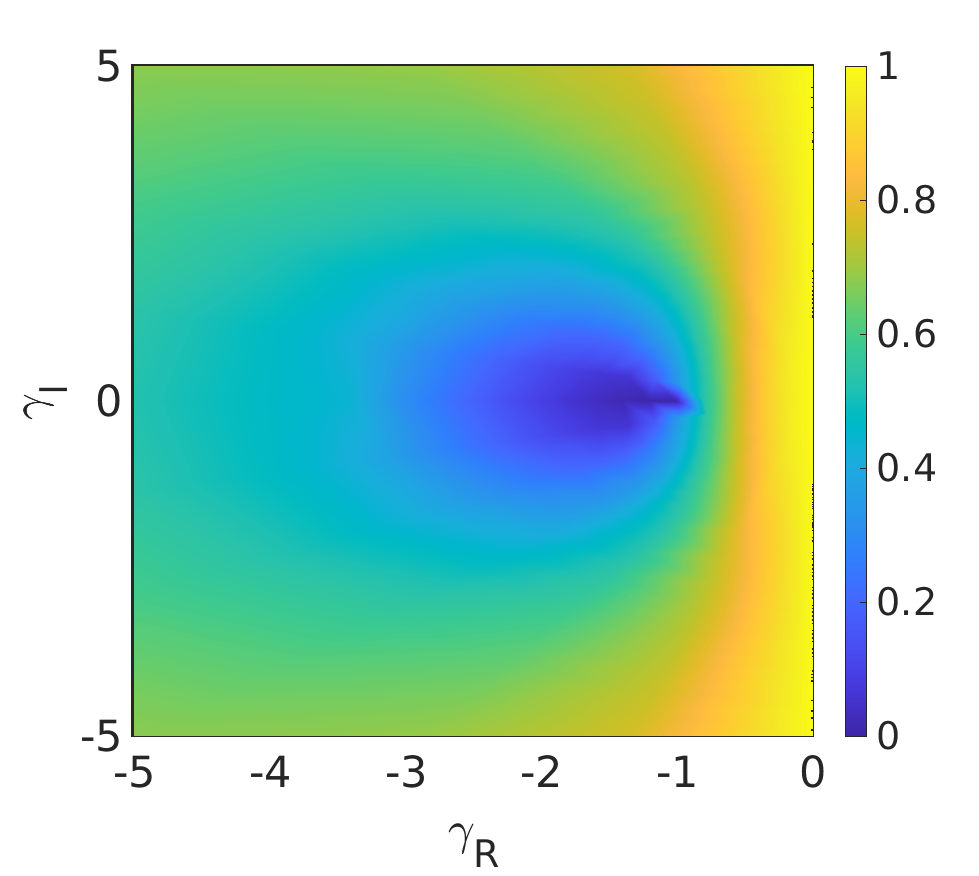}
    \caption{Localization ratio $|\Psi_3|^2/|\Psi_1|^2$ in the complex $\gamma$ plane. The coloring corresponds to this ratio. Note that ratio vanishes at $\gamma=-1$ where the singleton state stabilizes.}
    \label{fig:ZMphase}
\end{figure}

The presence of such a chiral symmetry suggests the existence of a symmetry protected zero mode analogous to the Su-Schrieffer-Heeger (SSH) model. The SSH model is the one-dimensional tight-binding model with alternating weak and strong hopping strengths, and its momentum-space topology is nontrivial when the hopping from the edge starts with a weak hopping. The topologically nontrivial SSH model hosts a zero energy edge state protected by the chiral symmetry whose wavefunction takes nonzero values in every other site. Our D$^4$ NLSE does not have an \textit{a priori} alternating hopping, however, since the density dependence affects the hopping strength, one can expect an emergent SSH-like physics induced dynamically by an appropriate density distribution. To explore such emergent SSH-like physics, we look for a zero energy edge mode. Such a zero energy edge mode must have nonzero wavefunction in every other site. We consider an open boundary condition with $j \ge 1$, and set the wave function on sites with even indices to zero $\Psi_{2n} = 0$. The D$^4$ NLSE taking zero chemical potential leads to the following condition to be satisfied by the wavefunctions in odd sites:
\begin{align}
   \left(J + \gamma |\Psi_{2n-1}|^2\right) \Psi_{2n-1} = \left(-J + \gamma^* |\Psi_{2n+1}|^2\right) \Psi_{2n+1}.
\end{align}
One can iteratively solve this equation for decreasing solution starting from some value of $\Psi_{1}$ for $\gamma_R < 0$ or $\Psi_{L}$ for $\gamma_R > 0$. When the total number of sites is odd, the iteratively-constructed solution is the exact zero energy solution of the D$^4$ NLSE. When the total number of sites is even, this construction fails as the final even site with only one adjacent odd site cannot properly satisfy the above equation. We note that there is a zero-energy mode completely localized at one site on the edge, i.e $n = 1$, when $\gamma/J = -1$, which we call a singleton state. (This state exists regardless of the number of sites). As $\gamma$ moves from $\gamma/J = -1$, the state becomes more and more delocalized as shown in Fig~\ref{fig:ZMphase}, which plots the ratio of the $|\Psi_3|^2/|\Psi_1|^2$ in the complex $\gamma$ plane as a measure of localization. The predominant trends show that as $|\gamma|$ moves from 1 and as $\gamma_I$ grows the state spreads further. In fact for a purely imaginary $\gamma$, the state is evenly spread on odd numbered sites.

\textit{Floquet Realization:---}
Finally, we present one method using Floquet engineering to realize the D$^4$ NLSE, which can be implemented in systems such as Bose-Einstein condensates of ultracold quantum gases and waveguide arrays.
We derive an effective Hamiltonian of a three step Floquet protocol as derived in the formalism presented in Ref.~\cite{Goldman14}.
We consider the time dependent quantum Hamiltonian 
\begin{align}
    H(t) &= -J\sum_j \hat{\Psi}^\dagger_{j+1}\hat{\Psi}_j + \hat{\Psi}^\dagger_j \hat{\Psi}_{j+1} + V(t)\\
    V(t) &= 
    \begin{cases}
        V_1, & 0 \le t < \frac{T_\omega}{3} \\
        V_2, & \frac{T_\omega}{3} \le t < \frac{2T_\omega}{3} \\
        -V_1 - V_2, & \frac{2T_\omega}{3} \le t < T_\omega
    \end{cases}\label{eqtn:3stepHN}
\end{align}
which corresponds to a free boson with pulsed modulation of the hopping parameter $V_1 = \sum_j J_1 \hat{\Psi}^\dagger_{j+1}\hat{\Psi}_j + \mathrm{H.c.}$ and a pulsed onsite interaction $V_2 = U\sum_j \hat{\Psi}^\dagger_j \hat{\Psi}^\dagger_j \hat{\Psi}_j \hat{\Psi}_j$ in three steps of equal duration with total period $T_\omega = 2\pi/\omega$. In the limit of large $\omega$ (i.e. $U/\omega, J_1/\omega \ll 1$), the effective Hamiltonian to first order in $1/\omega$ is given as
\begin{equation}
    \begin{split}
    H_{\mathrm{eff}} = \sum_j \hat{\Psi}^\dagger_{j+1}\left[-J + \frac{2i\pi UJ_1}{9\omega}(\hat{n}_j - \hat{n}_{j+1})\right]\hat{\Psi}_j\\
    +\hat{\Psi}^\dagger_{j}\bigg[-J - \frac{2i\pi UJ_1^*}{9\omega}(\hat{n}_j - \hat{n}_{j+1})\bigg]\hat{\Psi}_{j+1}
    \end{split}
\end{equation}
which maps directly onto Eq.~\ref{eq:H} with $\gamma = 2i\pi UJ_1/(27\omega)$ and identifying the quantum operators by classical variables assuming mean-field theory. We control the phase of the complex coupling $\gamma$ by changing the phase of the modulated hopping $J_1$. Complex hopping of $J_1$ can be engineered through modulation such as in Struck \textit{et al.} \cite{Struck12}, which can be combined with our proposed Floquet protocol, or through Raman-assisted tunneling \cite{JimenezGarcia12}. Similar Floquet interactions have been previously studied in various ultracold and optical systems~\cite{Centurion06,Keilmann11,Greschner14,Greschner15}.

\textit{Conclusion -} In this article we have presented the D$^4$NLSE and its phase diagram. We have derived the conditions for plane wave stability as well as a phase transition from the extended $k=0$ plane wave ground state to a localized ground state. Contrary to conventional NLSE, this model admits coexistence of a stable $k=0$ plane wave and a soliton ground state. We have further demonstrated that the interaction of the D$^4$ NLSE acts both either attractively or repulsively depending on the state in question. Although the D$^4$ NLSE itself is fully translationally invariant, the density distribution results in emergent SSH-like physics, and as a consequence, zero energy edge-localized mode exists. Finally we derived a possible method of Floquet engineering to realize the D$^4$ NLSE.

Many of the characteristic features of D$^4$ NLSE we found in this paper stem from the fact that the sign of the interaction is not determined \textit{a priori}, but rather dynamically determined by the density distribution of the nonlinear wave.  The emergent chiral symmetry analogous to the one discussed in this paper can be a versatile mechanism to obtain interaction-induced topology. 

The dynamical dependence of the sign of the interaction represents a new emerging paradigm in the field of nonlinear physics opening up unexpected and unexplored scenarios at both the classical and quantum levels. In this paper we have focused on  properties of  the classical ground state of the mean-field problem under this exotic form of dynamical gauge field. Ground-state phase diagram of the corresponding quantum many-body problems as well as its non-Hermitian generalizations are left for future studies.

\begin{acknowledgments}
\textit{Acknowledgments.---}
This work is supported by JSPS KAKENHI Grant No. JP20H01845, Grant No. JP21H01007, and JST CREST Grant No.JPMJCR19T1.
\end{acknowledgments}

\bibliography{biblio.bib}

\begin{widetext}

\appendix
\section{Continuum Model}
\label{app:ctnlim}
We now derive the continuum limit of our lattice model and study its properties.
Here we explicitly introduce the lattice spacing $\alpha$, and consider the limit $\alpha \to 0$ to find the coutinuous model. The Fourier space Hamiltonian of our model, with $\alpha$ explicitly written, is 
\begin{equation}
    H = \sum_k -2J\cos{(k\alpha)}\tilde{\Psi}_k\tilde{\Psi}_k + \frac{1}{L} \sum_{k,k',q} V(q\alpha,k\alpha) \tilde{\Psi}^*_{k} \tilde{\Psi}^*_{k'} \tilde{\Psi}_{k+k'-q} \tilde{\Psi}_{q} - \mu\sum_k \tilde{\Psi}^*_k \tilde{\Psi}_k,
    \label{eq:HlatTocnt}
\end{equation}
where 
\begin{align}
    V(q\alpha,k\alpha) = \gamma(e^{-iq\alpha} - e^{-ik\alpha}) - \gamma'(e^{iq\alpha} - e^{ik\alpha}).
\end{align}
In the continuum (long-wavelength) limit $k\alpha,q\alpha\rightarrow0$, the momentum dependent coefficients can be approximated through the Taylor expansion as
\begin{equation}
    \begin{split}
        -2J\cos{k\alpha} &\approx -2J + J(k\alpha)^2, \\
        V(q\alpha,k\alpha) &\approx 2i\gamma_R\alpha(k-q) + i\gamma_I\alpha^2(k^2 - q^2).
    \end{split}
\end{equation}
To obtain the continuum limit, we use these expansions in Eq.~\ref{eq:HlatTocnt} and perform the inverse Fourier transform
\begin{equation}
        \tilde{\Psi}_k = \frac{1}{\sqrt{L}}\sum_n \Psi_n e^{in\alpha k}.
\end{equation}
We define the continuum field operators at position $x = n\alpha$ by
\begin{align}
    \phi (x) = \Psi_n/\sqrt{\alpha},
\end{align}
with the identification $\int dx = \alpha \sum_n$.
The continuum limit of the kinetic and chemical potential parts of the Hamiltonian are then obtained as:
\begin{align}
        &\sum_k \left( -2J - \mu + J(k\alpha)^2 \right)\tilde{\Psi}^*_k\tilde{\Psi}_k
        =
        (-2J - \mu)\sum_n \Psi^*_n \Psi_n + 
        \frac{1}{L}\sum_k \sum_{n,m} J(k\alpha)^2 \Psi^*_n \Psi_m e^{ik\alpha (m-n)}\notag \\
        &=
        (-2J - \mu)\int dx \phi^*(x) \phi(x) + 
        \frac{\alpha}{L}\sum_k \int dx dx^\prime J k^2 \phi^*(x) \phi(x^\prime) e^{ik(x^\prime-x)}
        \notag \\
        &=
        (-2J - \mu)\int dx \phi^*(x) \phi(x) + 
        \frac{\alpha}{L}\sum_k \int dx dx^\prime J \phi^*(x) \phi(x^\prime) \frac{\partial^2}{\partial x \partial x^\prime}e^{ik(x^\prime-x)}
        \notag \\
        &=
        (-2J - \mu)\int dx \phi^*(x) \phi(x) + 
        \frac{\alpha}{L}\sum_k \int dx dx^\prime J \left( \partial_x \phi^*(x)\right) \left( \partial_{x^\prime }\phi(x^\prime)\right) e^{ik(x^\prime-x)}
        \notag \\
        &=
        (-2J - \mu)\int dx \phi^*(x) \phi(x) + 
        J \alpha^2 \int dx dx^\prime \left( \partial_x \phi^*(x)\right) \left( \partial_{x^\prime }\phi(x^\prime)\right) \delta (x - x^\prime)
        \notag \\
        &=
        (-2J - \mu)\int dx \phi(x)^* \phi(x) + 
        J \alpha^2 \int dx \left( \partial_x \phi^*(x)\right) \left( \partial_{x }\phi(x)\right).
\end{align}
Upon derivation, we have used integration by parts and ignored the boundary term (which is justified because we use the periodic boundary condition).
Similarly, we can derive the continuum limit of the interaction part of the Hamiltonian as:
\begin{equation}
        \frac{1}{L} \sum_{k,k',q} \bigg[2i\gamma_R\alpha(k-q) + i\gamma_I\alpha^2(k^2 - q^2)\bigg] \tilde{\Psi}^*_{k} \tilde{\Psi}^*_{k'} \tilde{\Psi}_{k+k'-q} \tilde{\Psi}_{q}.
\end{equation}
After some algebra, one can find that the term proportional to $\gamma_R$ is a total derivative and thus drops off. The continuum limit of the term proportional to $\gamma_I$ is
\begin{align}
    -i\gamma_I \alpha^3 \int dx \left[ \phi^* (x)^{2} \left( \partial_x \phi(x) \right)^2 - \left( \partial_x \phi^*(x) \right)^2 \phi (x)^2 \right].
\end{align}
Combining with the kinetic term, the total continuum Hamiltonian of the model is
\begin{align}
    H
    =
    \int dx \left[
    (-2J - \mu) \phi(x)^* \phi(x) + 
    J \alpha^2 \left( \partial_x \phi^*(x)\right) \left( \partial_{x }\phi(x)\right)
    -i \gamma_I \alpha^3 
    \left\{ \phi^* (x)^{2} \left( \partial_x \phi(x) \right)^2 - \left( \partial_x \phi^*(x) \right)^2 \phi (x)^2 \right\}
    \right]. \label{eq:hcont}
\end{align}
We can rewrite it in the following form to make it explicitly look as the density-gradient-dependent gauge field:
\begin{align}
    H =
    \int dx &\left[
    (-2J - \mu) \phi(x)^* \phi(x) + 
    J \alpha^2 \left\{ \left( i \partial_x + \frac{\gamma_I \alpha}{J}\partial_x |\phi(x)|^2 \right) \phi^*(x) \right\} \left\{ \left( -i\partial_{x} + \frac{\gamma_I \alpha}{J}\partial_x |\phi(x)|^2 \right) \phi(x)\right\} + \mathcal{O}(\gamma_I^2)
    \right],
\end{align}
where the density gradient term $\frac{\gamma_I \alpha}{J}\partial_x |\phi(x)|^2$ couples to the momentum $-i\partial_x$ analogous to the minimum coupling of the gauge field.

Let us now consider linearization around the zero momentum plane wave condensate state. Since the chemical potential of the plane wave condensate state is $\mu = -2J$, we use this value for the chemical potential, thereby dropping the first term in the Hamiltonian Eq.~(\ref{eq:hcont}) as: Firstly, we derive the equation of motions for the conjugate momenta $\pi$ and $\pi^*$ from the Hamilton equation $\dot{\pi} = \partial H / \partial (\phi) - \partial_x [ \partial H / \partial ( \partial \phi)]$ as:
\begin{equation}
    \begin{split}
        \dot{\pi} = -J \alpha^2 \partial_x^2\phi^* + 2i\gamma_I \alpha^3\bigg[\phi (\partial_x \phi^*)^2  +2\phi^*\partial_x\phi^*\partial_x\phi +{\phi^*}^2\partial_x^2\phi\bigg],\\
        \dot{\pi}^* = -J\alpha^2 \partial_x^2\phi - 2i\gamma_I \alpha^3 \bigg[\phi^* (\partial_x \phi)^2 + 2\phi\partial_x\phi^*\partial_x\phi + \phi^2\partial_x^2\phi^*\bigg].
    \end{split}
\end{equation}
Then we assume that the field takes the form of $\phi (x) = \frac{1}{\sqrt{L \alpha}} + \delta e^{ipx}$, which is a small plane-wave perturbation with momentum $p$ on top of a zero momentum plane wave state $1/\sqrt{L \alpha}$.
Keeping up to the linear order in $\delta$, we obtain
\begin{equation}
    \begin{pmatrix}
        \dot{\pi}\\
        \dot{\pi}^*
    \end{pmatrix} = 
    \begin{pmatrix}
        J \alpha^2 p^2 & -2i\gamma_I \alpha^2 p^2/L\\
        2i\gamma_I \alpha^2 p^2/L & J \alpha^2 p^2
    \end{pmatrix}
    \begin{pmatrix}
        \delta^*\\
        \delta
    \end{pmatrix}.
\end{equation}
The dispersion relation of the elementary excitation of the system is obtained by symplectic transformation of this equation, which is to diagonalize the two-by-two matrix in the above equation after multiplying $\sigma_z$ from the left.
The dispersion relation thus found is
\begin{equation}
    \pm (\alpha p)^2\sqrt{J^2 - \frac{4\gamma_I^2}{L^2}},
\end{equation}
which is exactly the same as the Bogoliubov excitation obtained in the main text in the long-wavelength limit. (Note that $p$ in the main text corresponds to $\alpha p$ here, writing out explicitly the lattice spacing $\alpha$.)

Finally, we show that the energy of this continuum model is not bounded from below as may be expected for a BEC with an attractive interaction. This can be seen assuming the following Gaussian ansatz wavefunction with variational parameters $a$ and $b$
\begin{equation}
    \phi(x) = \bigg(\frac{2a}{\pi}\bigg)^{1/4} e^{-(a+ib)x^2}
\end{equation}
Inserting this ansatz into the continuum Hamiltonian $H_\mathrm{cont}$, we find the energy of this state to be
\begin{equation}
    E(a,b) = J\alpha^2\frac{a^2+b^2}{a} + 2\gamma_I\alpha^3 \sqrt{\frac{a}{\pi}} b
    \label{eq:energyFtnl}
\end{equation}
In minimizing the energy, we obtain the following relation between our two parameters from setting the derivative with respect to $b$ to 0
\begin{equation}
    b = - \frac{\alpha\gamma_I}{J\pi^{1/2}}a^{3/2}.
\end{equation}
Plugging this back into Eq. \ref{eq:energyFtnl}, we find that the energy can be written as a function of $a$ as
\begin{equation}
    E = J\alpha^2a - \frac{\alpha^4\gamma_I^2}{J\pi} a^2
\end{equation}
from which one can immediately see that the energy goes to $-\infty$ as $a\rightarrow\infty$, namely by making the Gaussian narrower an narrower, and thus the energy of the continuum model is not bounded from below. This result implies that in the continuous case there is no analog of the soliton ground state of the lattice model. The localized soliton ground state is, therefore, a consequence of having a discrete lattice.

\section{Condensate Stability for Arbitrary $k$}
\label{app:arbk}
In this appendix we calculate the linear stability condition about an arbitrary momentum $k$ using the quantum Bogoliubov method. Starting from the quantum analogue of Eq.~\ref{eq:Hk}, we perform the Bogoliubov analysis about an arbitrary momentum $k$ to determine how the stability depends upon which momentum condensation occurs. We denote the quantum annihilation and creation operators corresponding to the classical operators $\tilde{\Psi}_p$ and $\tilde{\Psi}_p^*$ by $\tilde{a}_p$ and $\tilde{a}_p^\dagger$, respectively. The Hamiltonian is then 
\begin{equation}
    \begin{split}
    H = \sum_p -2J(\cos{p} - \cos{k})\tilde{a}^\dagger_p\tilde{a}_p + \frac{1}{L} \sum_{p,p',q} V(q,p) \tilde{a}^\dagger_{p} \tilde{a}^\dagger_{p'} \tilde{a}_{p+p'-q} \tilde{a}_{q}
    \end{split},
\end{equation}
where we have set the chemical potential to $\mu = -2J\cos{k}$ as is required for condensation at this momentum. For a condensate, we assume that $\tilde{a}^\dagger_k = \tilde{a}_k = \sqrt{N_k} \approx \sqrt{N}$ where $N_k$ is the number of particles of the $k$ momentum state and $N$ is the total number of particles. We linearize the quartic term by keeping only terms proportional to $N_k^2$ or $N_k$, i.e. interactions where at least two particles are in the momentum $k$ state. The quartic term thus contributes
\begin{align}
    \frac{N_k}{L}&\sum_p V(k,p+k)\tilde{a}^\dagger_{k+p}\tilde{a}^\dagger_{k-p} + V(p+k,k)\tilde{a}_{k-q}\tilde{a}_{k+q} \notag \\
    =&\frac{2iN_k}{L}\sum_p \bigg[\gamma_R\bigg(\sin(p+k) - \sin(k)\bigg) +\gamma_I\bigg(\cos(k) -\cos(k+p) \bigg) \bigg]\tilde{a}^\dagger_{k+p}\tilde{a}^\dagger_{k-p}\notag \\
    &- \bigg[\gamma_R\bigg(\sin(p+k) - \sin(k)\bigg) +\gamma_I\bigg(\cos(k) -\cos(k+p) \bigg) \bigg])\tilde{a}_{k-p}\tilde{a}_{k+p}\notag \\
    =&4in_k \bigg(\gamma_R\sin(k) - \gamma_I\cos(k)  \bigg)\sum_{p>0} \bigg[(\cos(p)-1) \bigg]\tilde{a}^\dagger_{k+p}\tilde{a}^\dagger_{k-p} - \bigg[ (\cos(p) -1) \bigg]\tilde{a}_{k+p}\tilde{a}_{k-p}\notag \\
    =&4in_k \gamma_k\sum_{p>0} \bigg[(\cos(p)-1) \bigg]\tilde{a}^\dagger_{k+p}\tilde{a}^\dagger_{k-p} - \bigg[ (\cos(p) -1) \bigg]\tilde{a}_{k+p}\tilde{a}_{k-p},
\end{align}
where we have used $V(p,k) = -V(k,p)$ to cancel the diagonal term and defined $\gamma_k = \gamma_R\sin{k} - \gamma_I\cos{k}$. The resulting Bogoliubov spectrum is
\begin{equation}
    \epsilon(p)^2 = 4J^2(\cos(p) - \cos(k))^2 - 16n_k^2\gamma_k^2(\cos(p)-1)^2,
\end{equation}
which results in the instability condition
\begin{equation}
    \frac{(\cos(p) - \cos(k))^2}{(\cos(p)-1)^2} < \frac{4n_k^2\gamma_k^2}{J^2}.
\end{equation}
The right hand side of this inequality will be zero when the condensation momentum $k$ satisfies the relation $\tan{k} = \gamma_I/\gamma_R$, and so condensation at this momentum will be stable while all other condensates will be unstable. Observe that the condition agrees with the calculation presented in the text for the $k=0$ condensate.

\end{widetext}

\end{document}